\documentclass[twocolumn,floatfix,aps,prl]{revtex4}
\usepackage{hyperref}
\hypersetup{breaklinks=true}
\hypersetup{colorlinks=true}
\hypersetup{urlcolor=blue}
\hypersetup{citecolor=blue}
\hypersetup{linkcolor=blue}
\begin{document}
\title{Comment on ``Excitation Spectrum and Superfluid Gap of an Ultracold Fermi Gas''}
\author{Yvan Castin}
\affiliation{Laboratoire Kastler Brossel, ENS-Universit\'e PSL, CNRS, Universit\'e Sorbonne and Coll\`ege de France, 24 rue Lhomond, 75231 Paris, France}
\maketitle

At zero temperature, as any three-dimensional superfluid with short-range interactions, a gas of paired fermionic atoms exhibits an acoustic excitation branch of low-wavenumber expansion $\omega_q=c q [1+\zeta q^2/k_{\rm F}^2+O(q^4\ln q)]$ with $c$ the speed of sound and $\zeta$ the curvature parameter (scaled by the Fermi wavenumber $k_{\rm F}$). Reference \cite{Biss} claims to have experimentally determined whether the branch has a convex $\zeta>0$ or concave $\zeta<0$ start, depending on the interaction strength. This is crucial information that dictates the nature of the gas relaxation mechanisms at low temperatures (the well studied three-phonon Beliaev-Landau or the yet unobserved four-phonon Landau-Khalatnikov mechanism \cite{LK,Annalen}). However, in this Comment, we argue that the high wavenumbers $q$ and temperature $T$ used in the experiment introduce a large bias capable of turning a convex branch into a concave one, so that further measurements at lower $q$ and $T$ are required to give a definitive answer.

To fix ideas, we consider the unitary limit $1/k_{\rm F}a=0$ with $a$ the $s$-wave scattering length, where the interactions are strongest and no solid argument can predict the sign of $\zeta$. Theoretically, one has $\zeta=-\pi^2$ $(2\xi_{\rm B})^{1/2}$ $[c_1+(3/2) c_2]$ where Bertsch's parameter gives the chemical potential in units of the Fermi energy $\mu=\xi_{\rm B} E_{\rm F}$ and $c_{1,2}$ quantify gradient corrections to quantum hydrodynamics \cite{SonWingate}. Only $\xi_{\rm B}$ is well known, $\xi_{\rm B}\simeq 3/8$ \cite{Zwierlein}. The dimensional expansion in powers of $\epsilon=4-d=1$ gives $c_1\simeq -0.0624(1-2\epsilon/3)+O(\epsilon^2)$ and $c_2=O(\epsilon^2)$ \cite{RupakSchafer}, so $\zeta>0$ to subleading order. Anderson's RPA, spectrally equivalent to the Gaussian fluctuations approximation of \cite{Randeria}, also predicts a positive value $\zeta_{\rm RPA}\simeq0.0838$ \cite{concav} (for $c_1\simeq-0.021$ \cite{RupakSchafer} this gives $c_2\simeq 0.0073\ll|c_1|$). The experimental value $\zeta_{\rm exp}=-0.085(8)$ \cite{Biss} is negative. However, assuming that the RPA is correct and that the branch start is convex, as we will do, actually has no clear incompatibility with the experiment, because the analysis in \cite{Biss} suffers from two serious limitations.

First, the value $\zeta_{\rm exp}$, obtained by cubic fitting of $\omega_q$ \cite{Biss}, could strongly depend on the fitting interval if too wide. In the RPA, fitting $\omega_q$ e.g. to the interval $0.22\leq q/k_{\rm F}\leq 1.08$ of Figure 1 in \cite{Biss} gives $\zeta_{\rm RPA}^{\rm fit}\simeq -0.026$, which even has the wrong sign. Since $\omega_q^{\rm RPA}$ has an inflection point at $q/k_{\rm F}\simeq 0.5$ \cite{concav} the fit blindly mixes convex and concave parts, which also explains the erroneous (negative) value of $\zeta_{\rm RPA}$ in \cite{Zwerger}.

Second, The high temperature $T\simeq 0.13 T_{\rm F}\simeq 0.8 T_{\rm c}$ ($T_{\rm c}$ is the superfluid transition temperature) in \cite{Biss} could modify the curvature of the acoustic branch by a non-negligible amount $\delta\zeta^{\phi\phi}$ via interaction with thermal phonons $\phi$. Treating the cubic phonon-phonon coupling $H_3^{\phi\phi}$ to second order and the quartic coupling $H_4^{\phi\phi}$ to first order, then taking the limit $k_{\rm B} T/mc^2\to 0$ ($m$ is the fermion mass), \cite{CRAS} obtains an expression for the thermal shift of $\omega_q$. This gives $\delta\zeta^{\phi\phi}\sim-[\pi^2/(3\xi_{\rm B})^{3/2}](T/T_{\rm F})^2$, or $\delta\zeta^{\phi\phi}\simeq -0.140$ at the experimental temperature. Since the small parameter used $k_{\rm B}T/mc^2\simeq 0.5$ is not $\ll 1$, we abandon the $T\to 0$ limit and add corrective curvature factors $(1\pm\alpha q^2/k_{\rm F}^2)$ to the amplitudes $\rho_q$ and $\phi_q$ of the superfluid density and phase quantum fluctuations ($\alpha=\pi^2(\xi_{\rm B}/2)^{1/2}[c_1-(3/2)c_2]\simeq-0.136$ \cite{Annalen}). We find $\delta\zeta^{\phi\phi}\simeq -0.110$, still negative enough to change the sign of curvature in the RPA. Furthermore, thermal pair dissociation creates fermionic quasiparticles $\gamma$ (another gas excitation branch) that interact with the phonons. Treating to second order the coupling $H_3^{\phi\gamma}$ and to first order the coupling $H_4^{\phi\gamma}$ given in \cite{PRL} with curvature factors in $\omega_q$, $\rho_q$ and $\phi_q$, we find $\delta\zeta^{\phi\gamma}\simeq -0.052$. This is a rough estimate: \cite{PRL} uses a simple local-density approximation whose small parameter $(k_{\rm B} T/m_*c^2)^{1/2}$ is $\approx1$ here (we use the $\gamma$ dispersion relation of \cite{NishidaSon} with the effective mass $m_*\simeq 0.56m$ and an energy minimum $\Delta=0.44 E_{\rm F}$ located at wavenumber $k_0=0.92 k_{\rm F}$ \cite{Ketterle}). Summing the thermal corrections gives $\zeta_{\rm RPA}^{\rm th}\simeq -0.078$ to compare with $\zeta_{\rm exp}=-0.085(8)$. 

The experiment, at first sight at odds with the RPA, could therefore very well be in agreement with it due to large finite-momentum and thermal bias, and the sign of curvature announced in \cite{Biss} may differ from the zero-temperature low-momentum one relevant for phonon damping.

{\sl Acknowledgments} -- We thank Alice Sinatra for her comments on the text.

\end{document}

\title{Commentaire sur ``Excitation Spectrum and Super\-fluid Gap of an Ultracold Fermi Gas''}
\author{Yvan Castin}
\affiliation{Laboratoire Kastler Brossel, ENS-Universit\'e PSL, CNRS, Universit\'e Sorbonne et Coll\`ege de France, 24 rue Lhomond, 75231 Paris, France}
\maketitle

Les auteurs de la référence \cite{Biss} mesurent entre autres la branche d'excitation acoustique d'un gaz superfluide d'atomes froids fermioniques appariés et déterminent, selon la force des interactions, si cette branche est concave ou convexe à faible nombre d'onde $q$. Ils franchissent ainsi une étape décisive dans l'étude expérimentale de ce système car ceci renseigne sur la nature des mécanismes de retour à l'équilibre du gaz à basse température, et pas seulement sur son état stationnaire. Il est donc important de savoir si les résultats annoncés sur la concavité de la branche acoustique sont fiables et définitifs, ou s'ils doivent être confirmés par des études ultérieures. Tel est l'objectif du présent commentaire. 

Rappelons qu'à température nulle, la branche acoustique du gaz de fermions appariés admet, comme dans tout système tridimensionnel ($d=3$) entièrement superfluide siège d'interactions à courte portée, le développe\-ment limité $\omega_q=c q [1+\zeta q^2/k_{\rm F}^2+ O(q^4\ln q)]$ où $\omega_q$ est la pulsation propre, $c$ la vitesse du son et $\zeta$ le paramètre de courbure (adimensionné par le nombre d'onde de Fermi $k_{\rm F}$). Le signe de $\zeta$ détermine, dans le régime faiblement collisionnel, le mécanisme d'amortissement des phonons $\phi$ à basse température. Si $\zeta>0$ (départ convexe), c'est celui à trois phonons de Belyaev et Landau, bien étudié dans différents systèmes. Si $\zeta<0$ (départ concave), c'est celui à quatre phonons de Landau et Khalatnikov, dont la théorie originale \cite{LK} a été rectifiée \cite{Annalen}, mais qui n'a encore été vu dans aucun système ; savoir que $\zeta<0$ dans le gaz de fermions est donc crucial pour une première observation.

Pour fixer les idées, restreignons-nous à la limite unitaire $1/k_{\rm F}a=0$ avec $a$ la longueur de diffusion dans l'onde $s$, où les interactions sont les plus fortes. Le pro\-blè\-me est alors non trivial et nous ne connaissons pas d'argument théorique solide fixant le signe de $\zeta$. 

Qualitativement, la relation de dispersion des quasi-particules fermioniques $\gamma$ (une autre branche d'excitation du gaz) atteint son minimum $\Delta\simeq 0,44 E_{\rm F}$ à un nombre d'onde non nul $k_0\simeq 0,92 k_{\rm F}$ \cite{Ketterle}, si bien que le continuum de paire brisée admet un bord plat de valeur $2\Delta$ jusqu'à $q=2k_0$ ($E_{\rm F}=\hbar^2k_{\rm F}^2/2m=k_{\rm B} T_{\rm F}$ est l'énergie de Fermi, $m$ la masse d'un atome) ; ce bord repousse la branche acoustique et tend à la rendre concave. Rien n'interdit cependant que la branche acoustique reste convexe (comme elle l'est naturellement dans la limite $1/k_{\rm F} a\to +\infty$) sur un voisinage de $q=0$, là où l'effet de répulsion par le continuum est le plus faible (l'effet de répulsion l'emporte partout dans la limite $1/k_{\rm F}a\to-\infty$). 

Quantitativement, on dispose de l'écriture $\zeta=-\pi^2$ $(2\xi_{\rm B})^{1/2}$ $[c_1+(3/2) c_2]$ où le paramètre de Bertsch donne le potentiel chimique $\mu=\xi_{\rm B} E_{\rm F}$ et les paramètres $c_{1,2}$ quantifient les corrections de gradient à l'hydrodynamique quantique \cite{SonWingate}. Seul $\xi_{\rm B}$ est bien connu, $\xi_{\rm B}\simeq 3/8$ \cite{Zwierlein}. Un développement dimensionnel en puissances de $\epsilon=4-d=1$ donne $c_1\simeq-0,0624[1-2\epsilon/3+O(\epsilon^2)]$ et $c_2=O(\epsilon^2)$ \cite{RupakSchafer} donc $\zeta>0$ à l'ordre sous-dominant. La RPA (Random Phase Approximation, c'est-à-dire approximation de la phase aléatoire) d'Anderson, spectralement équivalente à l'approximation des fluctuations gaussiennes de la référence \cite{Randeria}, prédit de manière non contrôlée une valeur positive, $\zeta_{\rm RPA}\simeq 0,0838$ \cite{concav}  (ceci correspond à $c_2\simeq 0,0073$ si l'on croit à la valeur $c_1\simeq-0,021$ donnée dans la référence \cite{RupakSchafer}, ce qui est effectivement $\ll |c_1|$). Mais la valeur expérimentale \cite{Biss} est de signe opposé, $\zeta_{\rm exp}=-0,085(8)$, donc ne conduit a priori pas au même mécanisme d'amortissement des phonons. L'enjeu est qualitatif et fondamental.

Cependant, comme nous allons le voir, on peut parfaitement supposer que la RPA soit correcte sans qu'il y ait d'incompatiblité nette avec la courbure négative vue dans l'expérience. L'analyse faite dans la référence \cite{Biss} souffre en effet de deux limitations. 

Premièrement, la valeur de $\zeta_{\rm exp}$ est obtenue par ajustement cubique de $\omega_q$ \cite{Biss}. Elle pourrait dépendre fortement de l'intervalle d'ajustement si celui-ci est trop étendu. Ainsi, dans la RPA, l'ajustement de $\omega_q$ sur l'intervalle $0,22\leq q/k_{\rm F}\leq 1,08$ de la figure 1 de la référence \cite{Biss} donnerait $\zeta_{\rm RPA}^{\rm fit}\simeq -0,026$, ce qui est faux, même en signe (en revanche, le biais serait $<2\%$ sur la vitesse du son $c$). En effet, $\omega_q^{\rm RPA}$ admet un point d'inflexion en $q/k_{\rm F}\simeq 0,5$ \cite{concav} si bien que l'ajustement mélange aveuglément des parties convexe et concave. Cet effet de mélange explique la valeur erronée ($<0$) de $\zeta_{\rm RPA}$ dans la référence \cite{Zwerger}.

Deuxièmement, la température assez élevée, $T\simeq 0,13 T_{\rm F}$ $\simeq 0,8 T_{\rm c}$ \cite{Biss} ($T_{\rm c}$ est la température de transition superfluide), pourrait modifier la courbure de la branche acoustique d'une quantité non négligeable $\delta\zeta^{\phi\phi}$ via l'interaction avec les phonons thermiques. En traitant le couplage phonon-phonon cubique $H_3^{\phi\phi}$ au second ordre et le couplage quartique $H_4^{\phi\phi}$ au premier ordre de la théorie des perturbations, puis en prenant la limite $k_{\rm B} T/mc^2\to 0$, la référence \cite{CRAS} obtient une expression analytique du déplacement thermique de $\omega_q$. Un développement à l'ordre $q^3$ donne alors l'équivalent exact 
\[
\delta\zeta^{\phi\phi}\sim-[\pi^2/(3\xi_{\rm B})^{3/2}](T/T_{\rm F})^2,
\]
soit $\delta\zeta^{\phi\phi}\simeq -0,140$ à la température expérimentale (la correction à la vitesse du son, en $O(T^4)$, est $\approx 1\%$). Comme le petit paramètre utilisé $k_{\rm B}T/mc^2\simeq 0,5$ n'est pas $\ll 1$, repartons du résultat perturbatif de la référence \cite{CRAS} sans prendre la limite $T\to 0$ et ajoutons des facteurs correctifs de courbure $(1\pm\alpha q^2/k_{\rm F}^2)$ aux amplitudes $\rho_q$ et $\phi_q$ des fluctations quantiques de densité et de phase du superfluide dans le mode de phonon de vecteur d'onde $\mathbf{q}$, avec $\alpha=\pi^2(\xi_{\rm B}/2)^{1/2}[c_1-(3/2)c_2]\simeq -0,136$ \cite{Annalen}. On trouve alors plutôt $\delta\zeta^{\phi\phi}\simeq -0,110$, ce qui reste assez négatif pour changer le signe de la courbure dans la RPA. De plus, la dissociation thermique des paires liées n'est pas négligeable dans l'expérience, le facteur de suppression $2\exp(-\Delta/k_{\rm B} T)$ valant quand même $0,07$. Or, les phonons interagissent avec les quasi-particules fermioniques qui en résultent. En traitant au second ordre le couplage cubique $H_3^{\phi\gamma}$ et au premier ordre le couplage quartique $H_4^{\phi\gamma}$ donnés dans la référence \cite{PRL} avec inclusion des facteurs de courbure dans $\omega_q$, $\rho_q$ et $\phi_q$, nous trouvons la contribution $\delta\zeta^{\phi\gamma}\simeq -0,052$. Cette estimation est grossière et incomplète : la référence \cite{PRL} utilise l'approximation d'homogénéité locale sans correction de gradient et son petit paramètre $(k_{\rm B} T/m_*c^2)^{1/2}$ (rapport des longueurs de cohérence de $\gamma$ et de $\phi$) est ici proche de un car la masse effective de $\gamma$ vaut $m_*\simeq 0,56m$ (nous utilisons la relation de dispersion de la référence \cite{NishidaSon}, avec un minimum d'énergie $\Delta=0,44 E_{\rm F}$ situé au nombre d'onde $k_0=0,92 k_{\rm F}$ comme il a été dit \cite{Ketterle}). La somme des corrections thermiques conduit à $\zeta_{\rm RPA}^{\rm th}\simeq -0,078$ à comparer à $\zeta_{\rm exp}=-0,085(8)$.

En résumé, le travail expérimental \cite{Biss}, aussi novateur soit-il, pourrait affirmer un peu vite avoir mesuré le coefficient de $q^3$ dans le développement limité de la branche acoustique $\omega_q$ du gaz unitaire de fermions à température nulle. Comme le suggèrent nos considérations simples, les valeurs élevées des nombres d'onde $q$ et de la température $T$ utilisées dans l'expérience introduisent un biais important capable de changer une branche acoustique convexe en une branche concave. Des études complémentaires sem\-blent requises. 

{\sl Remerciements -- } Nous remercions Alice Sinatra pour ses commentaires sur le texte.